\documentclass[sigconf]{acmart}

\usepackage{xspace}
\usepackage{booktabs}
\usepackage{caption}
\usepackage{subcaption}
\usepackage{enumitem}
\setlist[itemize]{leftmargin=*,label=\scalebox{.8}{\textbullet}}

\newcommand{\mathauthor}{\mathcal{A}}
\newcommand{\mathpaperset}{\mathcal{P}}
\newcommand{\mathpaper}{P}
\newcommand{\mathpersonas}{\mathrm{P}}
\newcommand{\specter}{\textsc{Specter}\xspace}
\newcommand{\preferencescore}{S\xspace}
\newcommand{\simTask}{\texttt{sT}\xspace}
\newcommand{\simTaskDistMethod}{\texttt{sTdM}\xspace}
\newcommand{\simspecter}{\texttt{ss}\xspace}

\AtBeginDocument{%
  \providecommand\BibTeX{{%
    \normalfont B\kern-0.5em{\scshape i\kern-0.25em b}\kern-0.8em\TeX}}}

\copyrightyear{2022}
\acmYear{2022}
\setcopyright{acmcopyright}\acmConference[CHI '22]{CHI Conference on Human Factors in Computing Systems}{April 29-May 5, 2022}{New Orleans, LA, USA}
\acmBooktitle{CHI Conference on Human Factors in Computing Systems (CHI '22), April 29-May 5, 2022, New Orleans, LA, USA}
\acmPrice{15.00}
\acmDOI{10.1145/3491102.3501905}
\acmISBN{978-1-4503-9157-3/22/04}

\begin{document}

%\title[Bridger: Toward Bursting Scientific Filter Bubbles]{Bridger: Toward Bursting Scientific Filter Bubbles\\ via Novel Author Discovery}

\title[Bursting Scientific Filter Bubbles]{Bursting Scientific Filter Bubbles: Boosting Innovation via Novel Author Discovery}


\author{Jason Portenoy}
\affiliation{%
 \institution{University of Washington}
 \country{}
}
\email{jporteno@uw.edu}

\author{Marissa Radensky}
\affiliation{
 \institution{University of Washington}
 \country{}
 }
\email{radensky@cs.washington.edu}

\author{Jevin West}
\affiliation{
 \institution{University of Washington}
 \country{}
 }
\email{jevinw@uw.edu}

\author{Eric Horvitz}
\affiliation{
 \institution{Microsoft}
 \country{}
 }
\email{horvitz@microsoft.com}

\author{Daniel S. Weld}
\affiliation{
\country{}
 \institution{Allen Institute for AI \\ University of Washington}
 }
\email{danw@allenai.org}

\author{Tom Hope}
\affiliation{
 \institution{Allen Institute for AI \\ University of Washington}
 \country{}
 }
\email{tomh@allenai.org}


\begin{abstract}
Isolated silos of scientific research and the growing challenge of information overload limit awareness across the literature and hinder innovation. Algorithmic curation and recommendation, which often prioritize relevance, can further reinforce these informational ``filter bubbles.'' In response, we describe Bridger, a system for facilitating discovery of scholars and their work. We construct a faceted representation of authors with information gleaned from their papers and inferred author personas, and use it to develop an approach that locates commonalities and contrasts between scientists to balance relevance and novelty. In studies with computer science researchers, this approach helps users discover authors considered useful for generating novel research directions. We also demonstrate an approach for \emph{displaying} information about authors, boosting the ability to understand the work of new, unfamiliar scholars. Our analysis reveals that Bridger connects authors who have different citation profiles and publish in different venues, raising the prospect of bridging diverse scientific communities.
\end{abstract}


\maketitle

\section{Introduction}\label{sec:introduction}

 \begin{quotation}
\noindent ``Opinion and behavior are more homogeneous within than between groups… Brokerage across structural holes provides a vision of options otherwise unseen.'' (Burt, 2004)
 \end{quotation}
 
The volume of papers in computer science continues to sky-rocket, with the DBLP computer science bibliography listing hundreds of thousands of publications in the year 2020 alone.
%\footnote{\url{https://dblp.org/statistics/publicationsperyear.html}} 
In particular, the field of AI has seen a meteoric growth in recent years, with new authors entering the field every hour \cite{tang2020pace}. Research scientists rely largely on search and recommendation services like Google Scholar and Semantic Scholar to keep pace with the growing literature and the authors who contribute to it. The literature retrieval services algorithmically decide what information to serve to scientists~\cite{beel2009google,cohan_specter_2020}, using information such as citations and textual content as well as behavioral traces such as clickthrough data, to inform machine learning models that output lists of ranked papers or authors.  
\begin{figure}[t]
    \centering
    \includegraphics[width=\columnwidth]{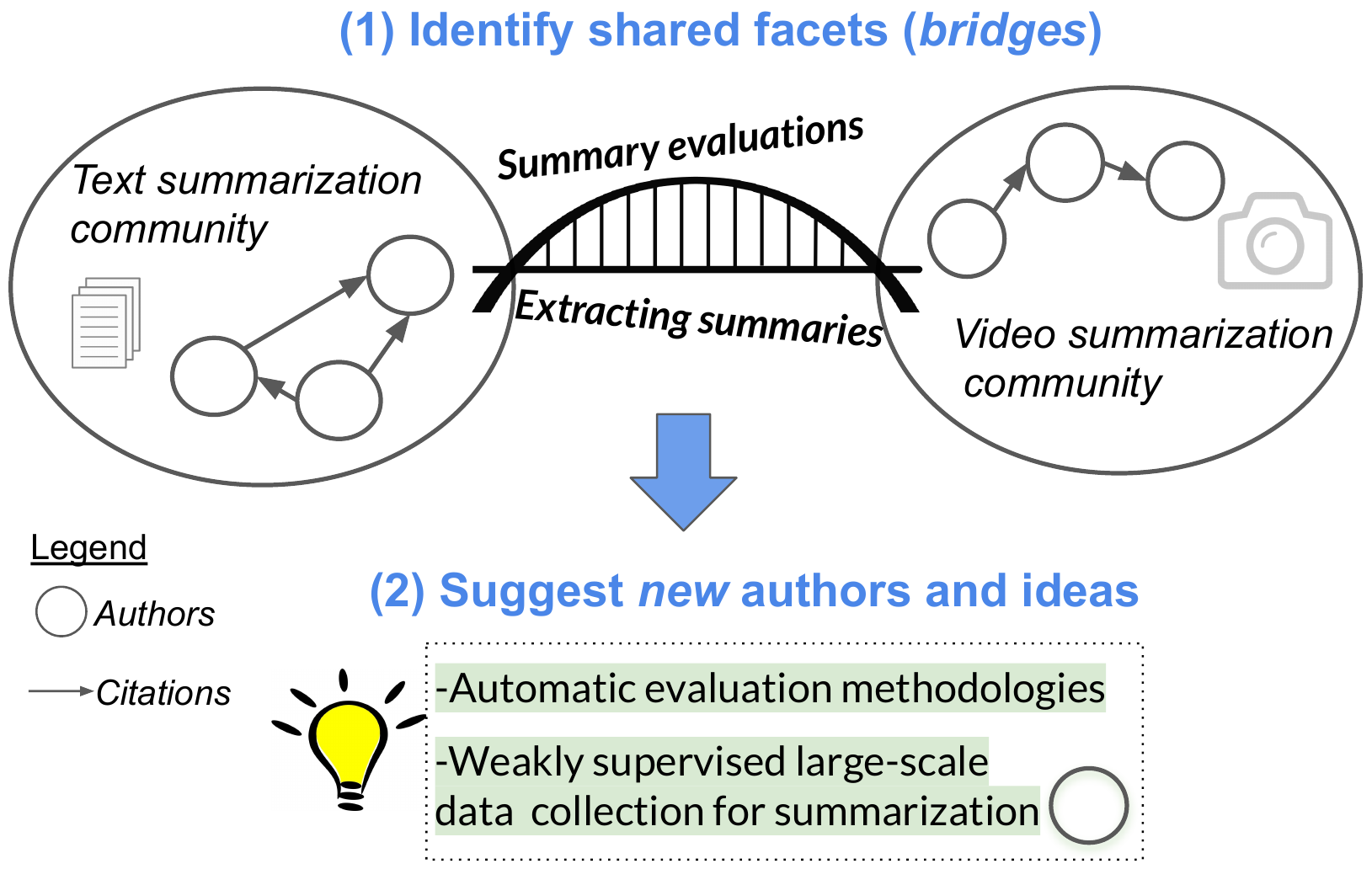}
    \caption{Our overarching goal is to (1) find commonalities among authors working in different areas and (2) suggest novel and valuable authors and their work, unlikely discovered otherwise due to their disparities. }
    \Description{This is a teaser figure which represents the Bridger system. The top is labeled "(1) Identify shared facets (bridges)." On the left is a bubble representing the "Text summarization community," with circles representing authors, and arrows between the circles to show that they are communicating amongst themselves, within-community. On the right is a similar bubble for "Video summarization community". Between the two bubbles is a bridge, labeled with two concepts that could bridge the two communities: "Summary evaluations," and "Extracting summaries". The bottom of the figure is labeled "Suggest new authors and ideas". Another circle is present to represent a new author, and two new suggested ideas are presented: "Automatic evaluation methodologies," and "Weakly supervised large-scale data collection for summarization".}
    \label{fig:teaser}
\end{figure}
By relying on user behavior and queries, these services adapt and reflect human input and, in turn, influence subsequent search behavior. This cycle of input, updating, engagement, and response can lead to an amplification of biases around searchers' prior awareness and knowledge \cite{Kim2016asa}. Such biases include selective exposure~\cite{frey1986recent}, homophily~\cite{mcpherson2001birds}, and the aversion to information from novel domains that require more cognitive effort to consider~\cite{hope_accelerating_2017,kittur_scaling_2019}. By reinforcing these tendencies, systems that filter and rank information run the risk of engendering so-called \emph{filter bubbles} \cite{pariser2011filter} that fail to show users novel content outside their narrower field of interest.

These bubbles and silos of information can be costly to individual researchers and for the evolution of science as a whole. They may lead scientists to concentrate on narrower niches~\cite{klinger2020narrowing}, reinforcing citation inequality and bias~\cite{nielsen2021global}, limiting cross-fertilization among different areas that could catalyze innovation~\cite{hope_accelerating_2017, kittur_scaling_2019, hope2021scaling}, and preventing knowledge brokerage across groups that has been associated with creativity and success \cite{burt_structural_2004}.
Addressing filter bubbles in general, in domains such as social media and e-commerce recommendations, is a hard and unsolved problem~\cite{ge2020understanding,chen2020improving,zhu2020measuring}. The problem is especially difficult in the scientific domain. The scientific literature consists of complex models and theories, specialized language, and an endless diversity of continuously emerging concepts. Connecting blindly across these cultural boundaries requires significant cognitive effort~\cite{vilhena_finding_2014}, translating to time and resources most researchers are unlikely to have at their disposal to enter unfamiliar research territory.\footnote{The challenge of limited time to explore novel directions is also discussed in our interviews with researchers; see §\ref{sec:exp2interview}.}

Our vision in this paper is to develop an approach that \textbf{boosts scientific innovation and builds bridges across scientific communities}, by helping scientists \textbf{discover authors that spark new ideas} for research directions. Working toward this goal, we developed Bridger, illustrated in Figure \ref{fig:teaser}. Our main contributions include: 

\noindent 
\begin{itemize}[topsep=1mm,noitemsep]

\item \textbf{A multidimensional author representation for matching authors along specific facets}. Our novel representation includes information extracted automatically from papers, specifically tasks, methods and resources, and automatically inferred \emph{personas} that reflect the different focus areas on which each scientist works. Each of these aspects is embedded in a vector space based on its content, allowing the system to \emph{identify authors with commonalities along specific dimensions} and not others, such as authors working on similar tasks but not using similar methods. 

\item \textbf{Boosting discovery of useful authors and ideas from novel areas.} We explore the utility of our author representation in experiments with computer science researchers interacting with Bridger. We find that this representation helps connect users with authors considered novel \emph{and} relevant, \emph{assisting users in finding potentially useful research directions}. Bridger outperforms a strong neural model currently employed by a public scholarly search engine for search and recommendation\footnote{\url{https://twitter.com/SemanticScholar/status/1267867735318355968}.}--- despite Bridger’s focus on surfacing \emph{novel} content and the built-in biases associated with this novelty. We conduct in-depth interviews with researchers, studying the tradeoffs between novelty and relevance in scientific content recommendations and discussing challenges and directions for author discovery systems.

\item \textbf{Exploring how to effectively depict recommended authors.} In addition to assessing {\em what} authors to recommend to spark new ideas for research directions, we also consider \emph{how} to display authors in a way that enables users to rapidly understand what new authors work on. We employ Bridger as an experimental platform to explore which facets should be displayed to users, investigating various design choices and tradeoffs. We obtain substantially better results in terms of user \emph{understanding of profiles of unknown authors}, when displaying information taken from our author representation. 

\item \textbf{Evidence of bridging across research communities.} Finally, we conduct in-depth analyses revealing that Bridger surfaces novel and valuable authors and their work that are unlikely to be discovered in the absence of Bridger due to publishing in different venues, citing and being cited by non-overlapping communities, and having greater distances in the social co-authorship network. 

\end{itemize}

Taken together, the ability to uncover novel and useful authors and ideas for research directions, and to serve this information to users in an effective and intuitive manner, suggests a future where automated systems are put to work to build bridges across communities, rather than blindly reinforcing existing filter bubbles.

\section{Related Work}\label{sec:related}

\paragraph{Inspirational Stimuli} Our work is related to literature focused on computational tools for boosting creativity \cite{hope_accelerating_2017,chan2018solvent,kittur_scaling_2019,goucher2020adaptive,hope2021scaling}.  Experiments in this area typically involve giving participants a specific concrete problem, and examining methods for helping them come up with creative solutions~\cite{hope_accelerating_2017,hope2021scaling}. In our efforts reported in this paper, we do not assume to be given a single concrete problem. Rather, we are given \emph{authors} and their papers, and automatically identify personalized inspirations in the form of other authors and their contributions. These computationally complex objects --- authors can have many papers with different themes, each paper with many facets and authored by multiple co-authors --- are very different to the short, single text snippets typically used in this line of work~\cite{hope_accelerating_2017,hope2021scaling}, or even to paper abstracts \cite{chan2018solvent}. A recurring theme in this area is the notion of a ``sweet spot'' for inspiration: not too similar to a given problem that a user aims to solve, and not too far afield~\cite{fu_meaning_2013}. Finding such a sweet spot remains an important challenge. Some work attempts to find this sweet spot by identifying  analogies as inspirations --- abstract structural relations between ideas \cite{hope_accelerating_2017,kittur_scaling_2019,hope2021scaling}. We study a related notion, balancing commonalities and contrasts between researchers for discovering authors that spark new research directions, trading off \emph{relevance} and \emph{novelty}. In our work, commonalities represent shared facets between authors (e.g., similar tasks) intended to help surface relevant authors, while contrasts along other dimensions (e.g., dissimilar methods) help promote novelty.

\noindent \paragraph{Filter Bubbles and Recommendations}
How to mitigate the filter bubble effect is a challenging open question for algorithmic recommendation systems~\cite{nguyen2014exploring}, explored recently for movies~\cite{zhu2020measuring} and in e-commerce \cite{ge2020understanding} by surfacing content that is aimed at being both novel and relevant. One approach that has been explored for mitigating these biases is judging recommendations not only by accuracy, but with other metrics such as diversity (difference between recommendations)~\cite{wilhelm2018practical,chen2020improving}, novelty (items assumed unknown to the user)~\cite{zhao2016much}, and serendipity (a measure of relevance and surprise associated with a positive emotional response)~\cite{wang2020impacts}. The notion of serendipity is notoriously hard to quantitatively define and measure~\cite{kaminskas2016diversity,chen2019serendipity,wang2020impacts, west2016IEEE}; recently, user studies have explored human perceptions of serendipity \cite{chen2019serendipity,wang2020impacts}, yet this problem remains very much open. A distinct, novel feature of our work is the focus on the scientific domain, and that unlike the standard recommendation system setting we measure our system's utility in terms of boosting users' ability to discover authors that spur \emph{new ideas for research directions}. In experiments with computer science researchers, we explore interventions that could potentially help provide bridges to authors working in diverse areas, with an approach based on finding faceted commonalities and contrasts between researchers. Our approach is broadly related to literature-based discovery \cite{swanson1996undiscovered}, where the goal is to surface scientific hypotheses by identifying potential links between concepts (e.g., drugs and diseases) that are not apparent by reading individual papers. Work in this area typically does not evaluate in the context of inspiring human users but rather in the ability to predict future links between biomedical entities (e.g., new links between drugs and diseases) \cite{nadkarni2021scientific}. Furthermore, in our work we focus explicitly on surfacing potential links between \emph{authors}, complex ``objects'' with many papers and lines of work. 

 \paragraph{Scientific Recommendations} Work in this area typically focuses on recommending \emph{papers}, using proxies such as citations or co-authorship links in place of ground truth~\cite{tang2012cross,beel2016paper,portenoy2020constructing,cohan_specter_2020}. In addition to being noisy proxies in terms of relevance, these signals reinforce existing patterns of citation or collaboration, and are not indicative of papers or authors that would help users generate \emph{novel} research directions --- the focus of Bridger. Furthermore, we perform controlled experiments with researchers to be able to better evaluate our approach without the biases involved in learning from observational data on citations or co-authorship. One related recent direction considers the problem of diversifying social connections made between academic conference attendees~\cite{tsai2018beyond,tsai2020diversity,wang2019sustainable}, by definition a relatively narrow group working in closely-related areas, using attendee metadata or publication similarity.
 
 \paragraph{Visualization-aided Exploration of the Scientific Literature}
There is a large body of work on the topic of mapping and visualizing networks of scholarly publications~\cite{borner_visualizing_2005,cobo_science_2011,chen_expert_2017}. Recently, attempts have been made in the information visualization research community to build tools for exploring connected aspects of the literature using \textit{interactive} visualizations~\cite{dork_pivotpaths:_2012,2015-refinery,heimerl_citerivers_2016,portenoy_leveraging_2017,hope2020scisight}. One recent paper~\cite{narechania_vitality_2021} designed a system to promote serendipitous discovery of new papers by finding semantically similar papers in a word embedding space; however, relying on such embeddings tuned for document similarity can reinforce filter bubbles, as we argue and demonstrate in this paper. In particular, we show that a state-of-art neural embedding model used by a popular scientific search engine for representing papers, underperforms our approach when it comes to discovering authors and ideas for research directions that are not only \emph{relevant}, but also \emph{novel} and more diverse. Finally, our work also studies novel design choices for displaying information on recommended authors, in a manner that increases users' ability to understand the work of scholars in unfamiliar areas.

\section{Bridger: Approach Overview}\label{methods}

\begin{figure*}[ht]
    \centering
    \includegraphics[width=\linewidth]{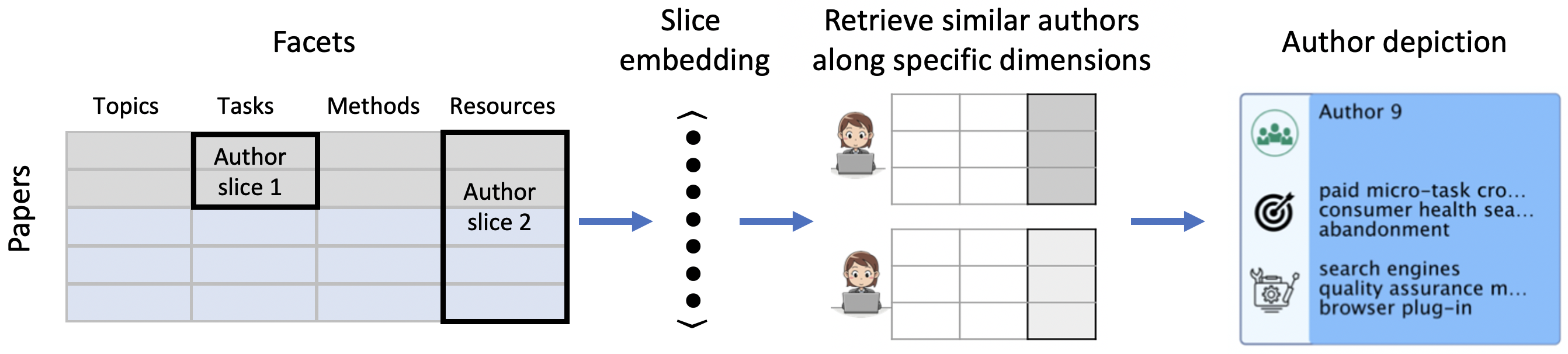}
    \caption{Overview of Bridger's author representation, retrieval, and depiction. Users are represented in terms of a matrix with rows corresponding to papers, and columns corresponding to facets. Bridger finds suggested authors who match along certain ``slices'' of the user's data -- certain facets, subsets of papers, or both.}
    \Description{The left of the figure shows the matrix representation of a user. Two example "slices" are shown. "Author slice 1" is a slice of two vertically adjacent cells, containing only the tasks taken from a single persona-based subset of the user's papers. "Author slice 2" is a slice of five vertically adjacent cells, representing only the resources from the user's total set of papers. To the right of the matrix, is a representation of the "slice embedding" which is a vector representation of one of these slices taken from the matrix. To the right of this, two new authors are shown, along with their own matrices. These are the new authors that the system has retrieved by finding similarities only along particular slices. Finally, on the far right, one of these new authors is shown represented as a card in the Bridger system, with that author's top tasks and methods listed.}
    \label{fig:bridger_overview}
\end{figure*}

In this section we present our novel faceted representation of authors, and methods for using this representation for author discovery by matching researchers along specific dimensions (Figure~\ref{fig:bridger_overview}). We also present methods for depicting the recommended authors when showing them to users. Bridger is designed to enable the study of different design choices for {connecting authors and ideas across scientific filter bubbles} and promoting discovery. We present the general framework, and the specific instantiations that we explore. 
We start by describing our representation for papers, and how Bridger represents authors by aggregating paper-level information and decomposing authors into {\em personas}.\footnote{The source code for data processing, author representation and ranking, and the user-facing application displaying this data, can be found in the supplementary materials.}

\subsection{Paper Representations}

Each paper $\mathpaper$ contains rich, potentially useful information. This includes raw text such as in a paper's abstract, incoming and out-going citations, publication date, venues, and more. One key representation we derive from each paper $\mathpaper$ is a vector representation $\tilde{\mathpaper}$, using a state-of-art scientific paper embedding model. This neural model captures overall coarse-grained topical information on papers, shown to be powerful in clustering and retrieving papers~\cite{cohan_specter_2020}.

Another key representation is based on fine-grained facets obtained from papers. Let $\mathcal{T}_{\mathpaper_i} =\{t_1, t_2,\ldots\}$ be a set of \emph{terms} appearing in paper $i$. Each term is associated with a specific \emph{facet} (category). Each term $t$ is located in a ``cell'' in the matrix illustrated in Figure \ref{fig:bridger_overview}, with facets corresponding to the columns and papers to rows. Each term $t \in \mathcal{T}_{\mathpaper_i}$ is also embedded in a vector space using a neural language model (see §\ref{subsec:implement}), yielding a $\tilde{t}$ vector for each term. 

We consider several categories of terms in this paper: coarse-grained paper topics inferred from the text \cite{wang_review_2019}, and fine-grained spans of text referring to \emph{methods, tasks and resources} automatically extracted from paper $i$ with a scientific named entity recognition model~\cite{wadden_entity_2019}. These three fine-grained categories are core aspects of computer science papers; in other words, they are key semantic concepts or ``building blocks'' with which computer scientists reason about research (developing new methods for a given task, developing new tasks, developing new datasets to support certain tasks, etc.)~\cite{luan_multi-task_2018,cattan2021scico}. These facets can help users find authors who spark ideas for new methods they can apply to their tasks, new tasks where their methods may be relevant, or new resources to explore. This relates to the fundamental role ``functional aspects'' play in science~\cite{hope_extracting_2021,hope2021scaling} and in linking between distant ideas and areas~\cite{chan2018solvent,hope_accelerating_2017,hope2021scaling}.

\subsection{Author Representations}
\label{subsec:authorrep}
We represent an author, $\mathauthor$, as a set of {\em personas} in which each persona is encoded with facet-wide aggregations of term embeddings across a set of papers.
Figure \ref{fig:bridger_overview} illustrates this with outlines of ``slices'' in bold --- subsets of rows and columns in the illustrated matrix, corresponding to personas (subsets of rows) and facets (columns).

\paragraph{Author Personas} Each author $\mathauthor$ can work in multiple areas. In our setting, this can be important for understanding the different interests of authors, enabling more control on author suggestions. We experiment with a clustering-based approach for constructing \emph{personas}, $\mathrm{P}_{\mathauthor}$, based on inferring for each set of author papers $\mathpaperset_{\mathauthor}$ a segmentation into $K$ subsets reflecting a common theme --- illustrated as subsets of rows in the matrix in Figure \ref{fig:bridger_overview}. We also experiment with a clustering based on the network of co-authorship collaborations in which $\mathauthor$ takes part. See §\ref{subsec:implement} for details on clustering. As discussed later (§\ref{sec:exp1}), we find that the former approach in which authors are represented with clusters of papers elicits considerably better feedback from scholars participating in our experiments.

\paragraph{Co-authorship Information} Each paper $\mathpaper$ is in practice authored by multiple people, i.e., it can belong to multiple authors $\mathauthor$. Each author assumes a \emph{position} $k$ for a given paper, potentially reflecting the strength of affinity to the paper. As discussed below (§\ref{subsec:implement}), we make use of this affinity in determining what weight to assign terms $\mathcal{T}_{\mathpaper_i}$ for a given paper and given author.

\paragraph{Author-level Facets} Finally, using the above information on authors and their papers, we construct multiple author-level \emph{facets} that capture different aggregate aspects of $\mathauthor$. More formally, in this paper we focus our experiments on author facets $\mathcal{V}_{\mathauthor} =\{\mathbf{m}, \mathbf{t}, \mathbf{r}\}$, where $\mathbf{m}$ is an aggregate embedding of $\mathauthor$'s \emph{method} facets, $\mathbf{t}$ is an embedding capturing  $\mathauthor$'s \emph{tasks}, and $\mathbf{r}$ represents $\mathauthor$'s \emph{resources}. In addition, we also construct these facets separately for each one of the author's personas $\mathrm{P}_{\mathauthor}$ --- corresponding to ``slice embeddings'' over subsets of rows and columns in the matrix illustrated in Figure \ref{fig:bridger_overview}. In analyses of our experimental results (§\ref{sec:exp2}), we also study other types of information such as citations and venues; we omit them from the formal notations to simplify presentation.

\subsection{Approaches for Recommending Authors}
\label{subsec:authorrec}
For a given author $\mathauthor$ using Bridger, we are interested in automatically suggesting new authors working on areas that are relevant to $\mathauthor$ but also likely to be interesting and spark new ideas for research directions. We are given a user $\mathauthor$, their set of personas $\mathpersonas_{\mathauthor}$, and for each persona its faceted representation $\mathcal{V}_{\mathauthor} =\{\mathbf{m}, \mathbf{t}, \mathbf{r}\}$.
We are also given a large pool of authors across computer science, $\{\mathauthor_1, \mathauthor_2,\ldots\}$, from which we aim to retrieve author suggestions to show $\mathauthor$. 

\paragraph{Baseline Model}We employ \specter, a strong neural model to which we compare, trained to capture overall topical similarity between papers based on text and citation signals (see \citet{cohan_specter_2020} for details) and used for serving recommendations as part of a large public academic search system. For each of author $\mathauthor$'s papers $\mathpaper$, we use this neural model to obtain an embedding $\tilde{\mathpaper}$. We then derive an aggregate author-level representation $\tilde{\mathbf{p}}$ (e.g., by weighted averaging that takes author-term affinity into account, see §\ref{subsec:implement}). Similar authors are computed using a simple distance measure over the dense embedding space. As discussed in the introduction and §\ref{sec:related}, this approach focuses on retrieving authors with the most overall similar papers to $\mathauthor$.
Intuitively, the baseline can be thought of as ``summing over'' both the rows and columns of the author matrix in Figure \ref{fig:bridger_overview}.
By aggregating across all of $\mathauthor$'s papers, information on finer-grained sub-interests may be lost. 
In addition, by being trained on citation signals, it may be further biased and prone to favor highly-cited papers or authors.

To address these issues, we explore a formulation of the author discovery problem in terms of matching authors along specific dimensions that allow more fine-grained control -- such as by using only a subset of views in $\mathcal{V}_{\mathauthor}$, or only a subset of $\mathauthor$'s papers, or both --- as in the row and column \emph{slices} seen in Figure~\ref{fig:bridger_overview}. This decomposition of authors also enables us to explore \emph{contrasts} along specific author dimensions, e.g., finding authors who use similar tasks to $\mathauthor$ but use very different methods or resources. 

\noindent \begin{itemize}
    \item \textbf{Single-facet matches:} For each author $\mathauthor_i$ in the pool of authors $\{\mathauthor_1, \mathauthor_2,\ldots\}$, we obtain their respective aggregate representations $\mathcal{V}_{\mathauthor_i} =\{\mathbf{m}, \mathbf{t}, \mathbf{r}\}$. We then retrieve authors with similar embeddings to $\mathauthor$ along one dimension (or matrix column in Figure \ref{fig:bridger_overview}; e.g., $\mathbf{r}$ for resources), ignoring the others. Unlike the baseline model, which aggregates \emph{all} information appearing in $\mathauthor$'s papers -- tasks, methods, resources, general topics, and any other textual information -- this approach is able to disentangle \emph{specific} aspects of an author, potentially enabling discovery of more novel, remote connections that can expose users to more diverse ideas and cross-fertilization opportunities.
    
    \item \textbf{Contrasts:} Finding matches along \emph{one} dimension does not guarantee retrieving authors who are \emph{distant} along the others. As an example, finding authors working on \emph{tasks} related to \textsl{scientific knowledge discovery} and \textsl{information extraction from texts}, could be authors who use a diverse range of \emph{resources}, such as \textsl{scientific papers}, \textsl{clinical notes}, etc. While the immense diversity in scientific literature makes it likely that focusing on similarity along one dimension only will still surface diverse results in terms of the other (see results in §\ref{sec:exp2}), we seek to further ensure this. 
    To do so, we apply a simple approach inspired by recent work on retrieving inspirations \cite{hope2021scaling}: We first retrieve the top $K$ authors $\{\mathauthor_1, \mathauthor_2,\ldots\, \mathauthor_K\}$ that are most similar to $\mathauthor$ along one dimension (e.g., $\mathbf{t}$), for some relatively \emph{large} $K$ (e.g., $K=1000$). We then rank this narrower list inversely by another dimension (e.g., $\mathbf{r}$), and show user $\mathauthor$ authors from the top of this list. Intuitively, this approach helps balance relevance and novelty by finding authors who are \emph{similar} enough along one dimension, and within that subset find authors who are relatively \emph{distant} along another.
    
    \item \textbf{Persona-based matching:} Finally, to account for the different focus areas authors may have, instead of aggregating over {\em all} of an author's papers, we perform the same single-view and contrast-based retrieval using the author's personas $\mathpersonas_{\mathauthor}$ --- or, in other words, row-and-column slices of the matrix in Figure \ref{fig:bridger_overview}. 
\end{itemize}

\subsection{Depicting Recommended Authors}
Our representation allows us to explore multiple design choices not only for {\em which} authors we show users, but also \emph{how} we show them. In our experiments (§\ref{sec:exp1}, §\ref{sec:exp2}), we evaluate authors' facets and personas in terms of their utility for helping researchers learn about new authors, and for controlling how authors are filtered.

\paragraph{Term Ranking Algorithms to Explain What Authors Work On} Researchers, flooded with constant streams of papers, typically have a very limited attention span to consider whether some new author or piece of information is relevant to them. It is thus important that the information we display for each author (such as their main methods, tasks, resources, and also papers) is \emph{ranked}, such that the most important or relevant terms appear first. We explore different approaches to rank the displayed terms, balancing between \emph{relevance} (or centrality) of each term for a given author, and \emph{coverage} over the various topics the author works on. We compare several approaches, including a customized relevance metric we design, in a user study with researchers (§\ref{sec:exp1}). We discuss in more detail the ranking approaches we try in §\ref{subsec:implement}.

\paragraph{Retrieval Explanations} In addition to term ranking approaches aimed at explaining to users of Bridger what a new suggested author works on, we also provide users with two rankings that are geared for explaining how the retrieved authors relate to them. First, we allow users to rank author terms $\mathcal{T}$ by how similar they are to their own list of terms (for each facet, separately). Second, users can also rank each author's \emph{papers} by how similar they are to their own --- showing the most similar papers first. These explanations can be regarded as a kind of ``anchor'' for increasing trust, which could be especially important when suggesting novel, unfamiliar content.

\subsection{Implementation Details}
\label{subsec:implement}

\subsubsection{Data}

We use data from the Microsoft Academic Graph (MAG)~\cite{sinha_overview_2015}. We use a snapshot of this dataset from March 1, 2021. We also link the papers in the dataset to those in a Semantic Scholar, a large public academic search engine.\footnote{\url{https://www.semanticscholar.org/}} We limit the papers and associated entities to those designated as Computer Science papers.
We focus on authors' recent work, limiting the papers to those published between 2015 and 2021, resulting in 4,650,474 papers from 6,433,064 authors. Despite using disambiguated MAG author data, we observe the challenge of author ambiguity still persists \cite{subramanian2021s2and}. In our experiments, we thus exclude participants with very few papers (see §\ref{sec:exp2}), since disambiguation errors in their papers stand out prominently.  

\subsubsection{Term Extraction}

We extract terms (spans of text) referring to tasks, methods, and resources mentioned in paper abstracts and titles, using the state-of-art DyGIE++ IE model \cite{wadden_entity_2019} trained on SciERC \cite{luan_multi-task_2018}.
We extracted 10,445,233 tasks, 20,705,854 methods, and 4,978,748 resources from 3,594,975 papers. We also use MAG topics, higher-level coarse-grained topics available for each paper in MAG. We expand abbreviations in the extracted terms using the algorithm in~\cite{schwartz_simple_2002} implemented in ScispaCy~\cite{neumann_scispacy_2019}.

\subsubsection{Scoring Papers by Relevance to an Author}\label{author-relevance-score}

The papers published by an author have varying levels of importance with regard to that author's overall body of publications. To capture this, we use a simple heuristic that takes into account two factors: the author's position in a paper as a measure of affinity (see §\ref{subsec:authorrep}), and the paper's overall impact in terms of citations. More formally, for
each author $\mathauthor$, we assign a weight $w_{\mathauthor,\mathpaper}$ to each paper $\mathpaper$ in $\mathpaper_{\mathauthor}$, $ w_{\mathauthor,\mathpaper} = \texttt{pos}_{\mathauthor,\mathpaper} \times \texttt{Rank}_{\mathpaper}$,
where $\texttt{pos}_{\mathauthor,\mathpaper}$ is 1.0 if $\mathauthor$ is first or last author on $\mathpaper$ and 0.75 otherwise,\footnote{This specific implementation reflects the norms around author position in computer science research. While many fields share these same norms, they are not universal, and so these methods can be adjusted when this system is applied to fields with different conventions.} and $\texttt{Rank}_{\mathpaper}$ is MAG's assigned paper Rank (a citation-based measure of importance, see~\cite{wang_review_2019} for details), normalized by min-max scaling to a value between .5 and~1.

\subsubsection{Author Similarity}

We explore several approaches for author similarity and retrieval, all based on paper-level aggregation as discussed in §\ref{subsec:authorrec}.
For the document-level \specter baseline model discussed in §\ref{subsec:authorrec}, we obtain 768-dimensional embeddings for all of the papers. To determine similarity between authors, we take the average embedding of each author's papers, weighted by the paper relevance score described above. We then compute the cosine similarity between this author and the average embedding of every other author.
For our faceted approach, we compute similarities along each author's facets, using embeddings we create for each term in each facet. The model used to create embeddings was CS-RoBERTa~\cite{gururangan_dont_2020}, which we fine-tuned for the task of semantic similarity using the Sentence-BERT framework~\cite{reimers_sentence-bert_2019}. For each author or persona, we calculate an aggregate representation along each facet by taking the average embedding of the terms in all of the papers, weighted by the relevance score of each associated paper. 

\subsubsection{Identification of Personas}

We infer author personas using two different approaches. For the first approach we cluster the co-authorship network using the ego-splitting framework in~\cite{epasto_ego-splitting_2017}.
In a second approach, we cluster each authors' papers by their \specter embeddings using agglomerative clustering with Ward linkage \cite{murtagh2014ward} on the Euclidean distances between embedding vectors.\footnote{Implemented in the scikit-learn Python library \cite{pedregosa2011scikit}. Distance threshold of 85.} In our user studies, we show participants their personas and the details of each one (papers, facets, etc.).\footnote{Some authors do not have detected personas; we observe this to often be the case with early-career researchers.} To make this manageable, we sort the clusters (personas) based on each cluster's most highly ranked paper according to MAG's assigned rank, and show participants only their top two personas.

\subsubsection{Term Ranking for Author Depiction} 

We evaluate several different strategies to rank terms (methods, tasks, resources) shown to users in Experiment I (§\ref{sec:exp1}):

\begin{itemize}
    \item \textbf{TextRank:} For each term $t$ in an author's set of papers, we create a graph $G_F = (V, E)$ with vertices $V$ the terms and weighted edges $E$, where weight $w_{ij}$ is the euclidean distance between the embedding vectors $\tilde{t_i}$ and $\tilde{t_j}$. We score each term $t_i$ according to its PageRank value in $G_F$~\cite{mihalcea_textrank_2004}.
    
    \item \textbf{TF-IDF:} For each $t$, we compute TF-IDF across all authors, considering each author as a ``document'' (bag of terms) in the IDF (inverse document frequency) term, counting each term once per paper. We calculate the TF-IDF score for each term for each author, and use this as the term's score.
    
    \item \textbf{Author relevance score:} For each $t$, we calculate the sum of the term's relevance scores (§~\ref{author-relevance-score}) derived from their associated papers. If a term is used in multiple papers, the associated paper's score is used for each summand.
    
    \item \textbf{Random:} Each term $t$ is assigned a random rank.
\end{itemize}

\section{Experiment I: Author Depiction}
\label{sec:exp1}

In systems that help people find authors, such as Microsoft Academic Graph, Google Scholar, and AMiner \cite{wan2019aminer}, authors are often described in terms of a few high-level topics. 
In advance of exploring how we might leverage facets to engage researchers with a diverse set of authors,
we performed a user study to gain a better understanding of what information might prove useful when depicting authors. We started from a base of Microsoft Academic Graph (MAG) topics, and then added their extracted facets (tasks, methods, resources). We investigated the following research questions:
\begin{itemize}
\item{\textbf{RQ1}: Do tasks, methods, and/or resources complement MAG topics in depicting an author’s research?}
\item{\textbf{RQ2}: Which term ranking best reflects an author's interests?}
\item{\textbf{RQ3}: Do tasks, methods, and/or resources complement MAG topics in helping users gain a better picture of the research interests of \emph{unknown} authors?}
\item{\textbf{RQ4}: Do personas well-reflect authors' different focus areas?}
\end{itemize}

\subsection{Experiment Design}
{After the experiment was approved for IRB exemption,} thirteen computer-science researchers were recruited for the experiment through Slack channels and mailing lists. Participants were compensated \$20 over PayPal for their time. Study sessions were one-hour, semi-structured interviews recorded over Zoom. The participants engaged in think-aloud throughout the study. They evaluated a depiction of a known author (e.g., research mentor) for accuracy in depicting their research, as well as depictions of five \emph{unknown} authors for usefulness in learning about new authors. 

Throughout all parts of the experiment, the interviewer asked follow-up questions regarding the participant's think-aloud and reactions.\footnote{The script for Experiment I can be found in our supplementary materials.} To address \textbf{RQ1} and \textbf{RQ2}, the participants first evaluated the accuracy of a known author's depiction.

\emph{Step I.} To begin, we presented the participant with only the top 10 MAG topics for the known author. We asked them to mark any topic that was unclear, too generic, or did not reflect the author's research well. Next, we provided five more potential lists of terms. One of these lists consisted of the next 10 top topics. The other four presented 10 tasks, each selected as the top-10 ranked terms using the strategies described in §\ref{subsec:implement}.
We asked participants to rank the five lists (as a whole) in terms of how well they complemented the first list (with an option to select none). 

\emph{Step II.} The process then repeated for five more potential lists to complement the original topics and the highest-ranked second list selected in Step I --- this time, with methods instead of tasks. %(and a fifth list of the next ten MAG topics). 
If the participant ranked a methods list highest, we then presented the participant with a resources list that used the same ranking strategy preferred by the participant for methods, and asked whether or not this list complemented those shown so far.  

\emph{Step III.} To address \textbf{RQ3}, participants next evaluated the utility of author depictions for five unknown authors. To describe each unknown author, we provided topics, tasks, methods, and resources lists with 10 terms each. The non-topics lists were ranked using TF-IDF as a default. The participant noted whether or not each additional non-topics list complemented the preceding lists in helping them understand what kind of research the unknown author does. 

\emph{Step IV.} Finally, for \textbf{RQ4}, we asked participants to evaluate the known author's distinct personas presented in terms of tasks, which were ranked using TF-IDF.  On a Likert-type scale of 1-5, participants rated their agreement with the statement, “The personas reflect the author's different research interests (since the year 2015) well.”

\subsection{Results}

{The think-aloud results were evaluated using thematic analysis \cite{braun2012thematic}. The transcripts of each participant were reviewed and paraphrased {with quotes and contextual} notes. Codes and themes were generated from the review. The notes for each participant (and full transcripts as necessary) were then reviewed again, and relevant codes were connected to each participant.}

\subsubsection{Results for RQ1}\label{sec:exp1:rq1}
\textbf{The majority of participants found that tasks, methods, and resources complemented topics to describe a known author's research.}
For both tasks and methods, 11 of 13 participants felt that seeing information about that facet, more so than additional top MAG topics or no additional information, complemented the original top ten MAG topics. The prevailing grievance with the additional MAG topics was that they were too general. {For example, looking at the topics column that was an alternative to the potential tasks columns, P7 said, "\emph{Some of it like `healthcare' and `applied psychology' are too high-level}." In the same situation, P2 commented, "\emph{AI is very general and implied by [the author's other topics] machine learning or NLP or IR}."} Furthermore, 7 of 9 participants who evaluated a resources list thought that it complemented the preceding lists.

\subsubsection{Results for RQ2}\label{sec:exp1:rq2}
\begin{figure}[tb]
  \centering
  \includegraphics[width=0.85\columnwidth]{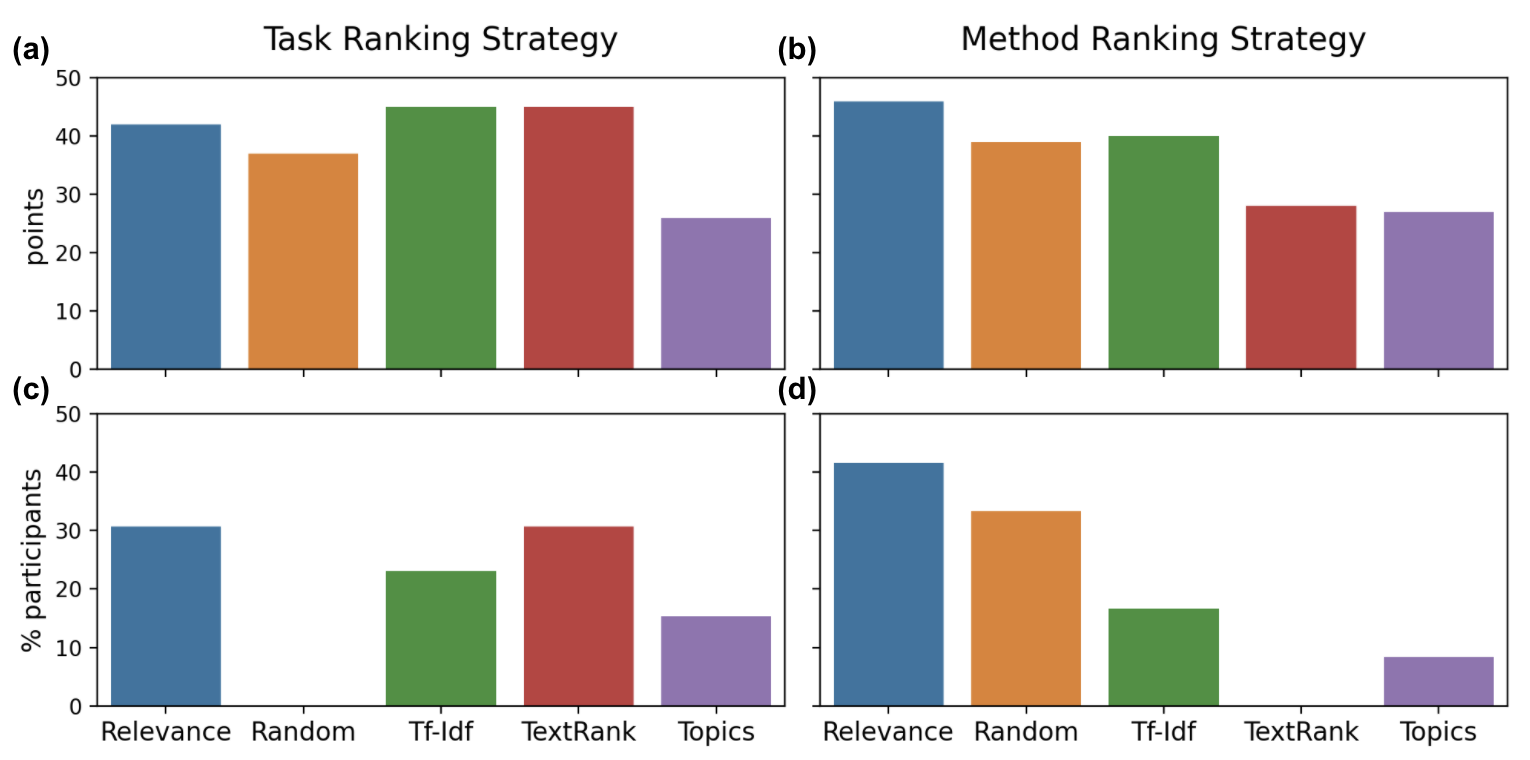}
  \caption{Points awarded to each ranking strategy for tasks (a) and methods (b), and percentage of participants who favored each strategy most for tasks (c) and methods (d).}
  \label{fig:accuracyStudy1}
  \Description{(a): The bar chart for total points awarded to each ranking strategy for tasks shows that TF-IDF and TextRank perform best with 45 points, and the relevance score trails close behind with 42 points. Random is next with 37 points, and the Topics baseline has just 26 points. (b): The bar chart for total points awarded to each ranking strategy for methods shows that the relevance score performs best with 46 points. TF-IDF follows with 40 points, and Random with 39 points. TextRank and Topics have 28 and 27 points respectively. (c): Participants selected the relevance score and TextRank most often as their favorite ranking strategy for tasks. Around 31 percent of participants favored each most. Next was TF-IDF with around 23 percent of votes, and then Topics with around 15 percent of votes. No one favored Random most. (d): Participants selected the relevance score most often as their favorite ranking strategy for methods. It received around 42 percent of votes. Next was Random with around 33 percent of votes, followed by TF-IDF with around 17 percent of votes. Topics received just around 8 percent of votes, and TextRank none.}
\end{figure}

\textbf{Participants overall preferred the relevance score ranking strategy for tasks and methods.} We compared the four ranking strategies and MAG topics baseline strategy for both tasks and methods. For each participant, we awarded points to each strategy based on its position in the participant's ranking of the five strategies (Figure~\ref{fig:accuracyStudy1}a, b). We awarded the least favorite strategy one point and the most favorite strategy five points. Since there were 13 participants, a strategy could accumulate up to 65 points. Separately, we counted how many times each strategy was a participant's favorite strategy (Figure~\ref{fig:accuracyStudy1}c, d). With regards to tasks, TextRank and TF-IDF accrued the most points from participants, with the relevance score trailing close behind (Figure \ref{fig:accuracyStudy1}a). Meanwhile, the MAG topics baseline accrued the least points, even fewer than the random task ranking strategy. In addition, relevance score and TextRank were chosen most often as the favorite task ranking strategy (Figure \ref{fig:accuracyStudy1}c). With regards to methods, the relevance score ranking strategy performed best in terms of both total points (Figure \ref{fig:accuracyStudy1}b) and favorite strategy (Figure \ref{fig:accuracyStudy1}d). 

\subsubsection{Results for RQ3}\label{sec:exp1:rq3}
\textbf{Participants generally found tasks, methods, and resources helpful to better understand what kind of research an unknown author does.} To calculate how many participants were in favor of including tasks, methods, and resources to help them better understand an author, we determined the average of each participant's binary response per facet. Adding up the 13 responses for each facet, we saw that the majority of participants thought each additional facet helped them understand the unknown author better. All 13 participants found the tasks helpful, eight found the methods helpful, and 12 found the resources helpful. As an example, P12 connected an unknown author's topics, tasks, and methods to better understand them: “\emph{I wouldn't have known they were an information retrieval person from the [topics] at all.... The previous things [in topics and tasks] that mentioned translation and information retrieval and kind of separately… This [methods section] connects the dots for me, which is nice.}” Interestingly, methods were not viewed to be as useful as tasks or resources. The majority of participants cited unfamiliar terms as a key issue. For instance, {after looking at the first four methods such as ``Experience Sampling Method'' and ``TapSense'' in an unknown author's methods column, P9 looked at the fifth one and noted, "\emph{'Body posture calculations'- I think that's the first phrase that I can pick out maybe what it means}."}

\subsubsection{Results for RQ4}\label{sec:exp1:rq4}

\textbf{Participants indicate preference for personas selected based on papers rather than co-authorship.} After the experiment, six participants were informally asked to compare the experiment's personas selected based on co-authorship with the personas based on paper-based clustering (see §\ref{subsec:implement}). Four of them preferred the updated version. Furthermore, one of the users who preferred the old version still thought the updated version had better personas themselves and merely did not like the updated personas' ordering. In addition, all six participants liked seeing the personas in terms of papers. In our experiment in §\ref{sec:exp2}, we observed much higher satisfaction with the updated personas in comparison to the original personas of this experiment.

\label{sec:qualResultsStudy1}

\section{Experiment II: Author Discovery}\label{sec:exp2}

We now turn to our main experiment, exploring whether facets can be employed in Bridger to spur users to discover valuable and novel authors and their work. We use our two author-ranking strategies (§\ref{subsec:authorrec}), one based on similar tasks alone (\simTask) and the other on similar tasks with contrasting (distant) methods (\simTaskDistMethod). We compare these strategies to the \specter (\simspecter) baseline. More specifically, we investigated the following research questions: 
\begin{itemize}
    \item{\textbf{RQ5}: Do \simTask and \simTaskDistMethod, in comparison to \specter, surface suggestions of authors that are considered novel and valuable, coming from research communities more distant to the user?}
    \item{\textbf{RQ6}: Does sorting based on personas help users find more novel and valuable author suggestions?}
\end{itemize}

\subsection{Experiment Design}\label{sec:exp2-design}

\begin{figure}[htb]
  \includegraphics[width=0.75\columnwidth]{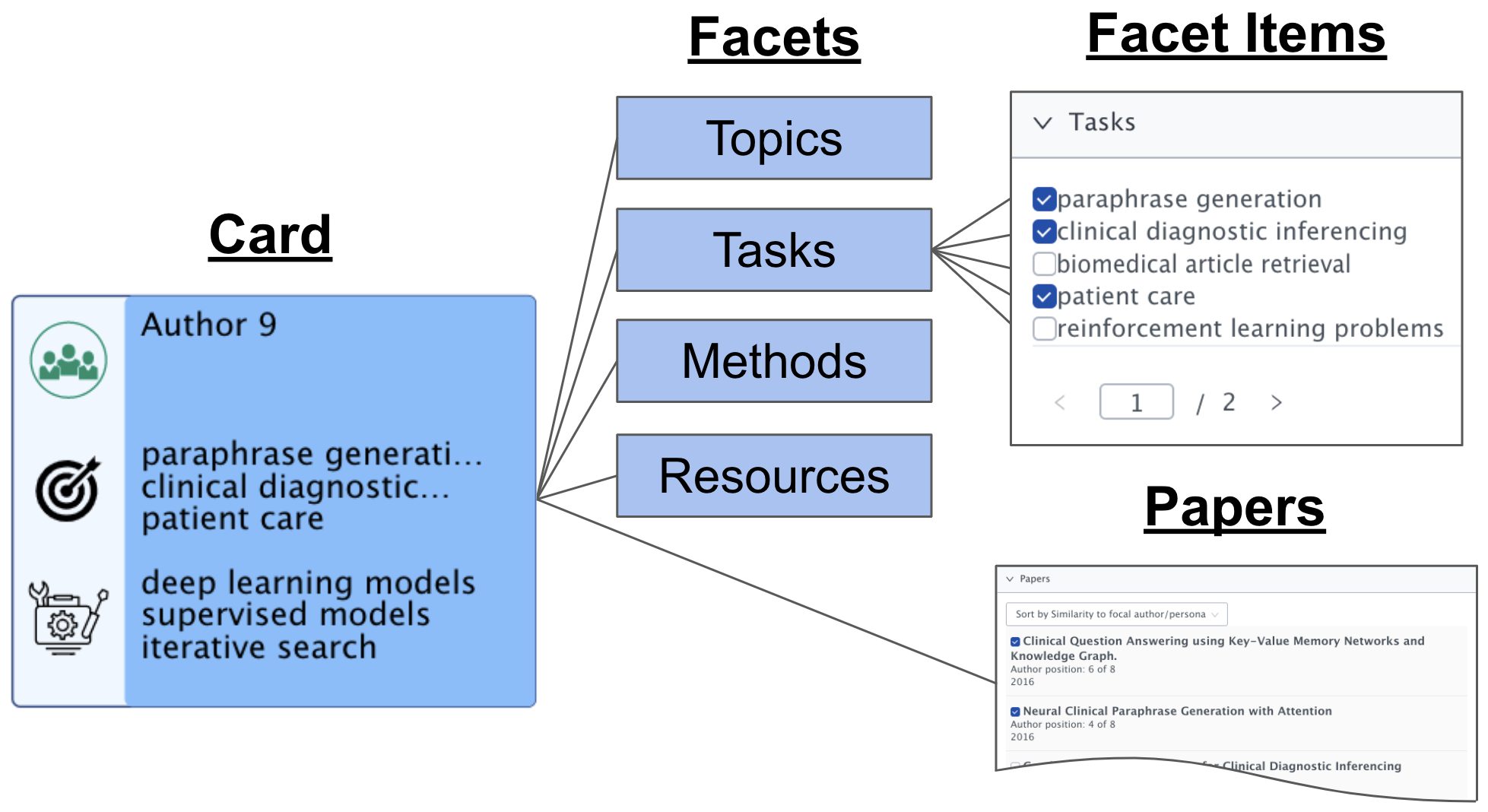}
  \caption{Illustration of information shown to users in Experiment II, §\ref{sec:exp2}. When the user clicks on an author card, an expanded view is displayed with 5 sections: papers, topics, and our extracted facets --- tasks, methods, and resources.}
  \label{fig:hierarchy}
  \Description{An author card is shown broken down into its five sections for the five facets: papers, topics, tasks, methods, and resources. Examples of the papers section and tasks section are displayed with a few papers and tasks checked off. In the papers section, there is a sort drop-down menu at the top that currently says "sort by similarity to focal/author persona." There are a couple papers listed with their titles, author position, and year of publication. In the tasks section, there are five tasks listed: paraphrase generation, clinical diagnostic inferencing, biomedical article retrieval, patient care, and reinforcement learning problems. At the bottom of the section, there are arrows to go back and forth between pages one and two of the task phrases.}
\end{figure}

Twenty computer-science researchers participated in the experiment {approved for IRB exemption}, after recruitment through
Slack channels and mailing lists. Participants were compensated \$50 over PayPal for their time.

All participants were shown results based on their overall papers (without personas) consisting of 12 author cards they evaluated one by one.
Four cards were included for each of \simTask, \simTaskDistMethod, and \simspecter. We only show cards for authors who are at least 2 hops away in the co-authorship graph from the user, filtering authors with whom they had previously worked.

For participants who had at least two associated personas, we also presented them with authors suggested based on each separate persona: four author cards for each of their top two personas (two under \simTask and two under \simTaskDistMethod). Whether the participants saw the personas before or after the non-persona part was randomized. 

Each author card provides a detailed depiction of that author (see Figure~\ref{fig:bridger_overview}). The author's name and affiliation is hidden in this experiment to mitigate bias. As shown in Figure \ref{fig:hierarchy}, cards showcase five sections of the author's research: their papers, MAG topics, and our extracted facet terms (i.e., tasks, methods, and resources). We also let users view the tasks and methods ranked by \emph{similarity} to them, which could be helpful to explain why an author was selected and better understand commonalities. 

The cards showed up to five items for each section, with some sections having a second page, depending upon data availability, for a maximum of ten items per section. 
Papers could be sorted based on recency or similarity to a  participant /  persona.
To avoid biasing participants, the only information provided for each paper was its title, date, and the suggested author's position on each paper (e.g., first, last).

Each of these items (papers and terms) had a checkbox, which the user was instructed to check if it fulfilled two criteria: 1) potentially interesting and valuable for them to learn about or consider in terms of utility, and 2) not too similar to things they had worked on or used previously. Following a short tutorial,\footnote{The tutorial slides are available in our supplementary materials.} participants evaluated each author shown by checking the aforementioned checkboxes (see Figure~\ref{fig:hierarchy}, right). While evaluating the first and last author (randomized), the participant engaged in a protocol analysis methodology (sharing their thinking as they worked). Participants with personas were also asked, based on each persona's top five associated papers, whether they each reflected a coherent focus area, and whether they seemed useful for filtering author suggestions.\footnote{See supplementary materials for the source code used for generating the data for Experiment II, as well as the code for the interactive application used in the evaluation, and the script used to direct the participants.}

\subsection{Quantitative Results}\label{sec:exp2-quant-results}

\begin{figure*}
  \centering
  \includegraphics[width=\textwidth]{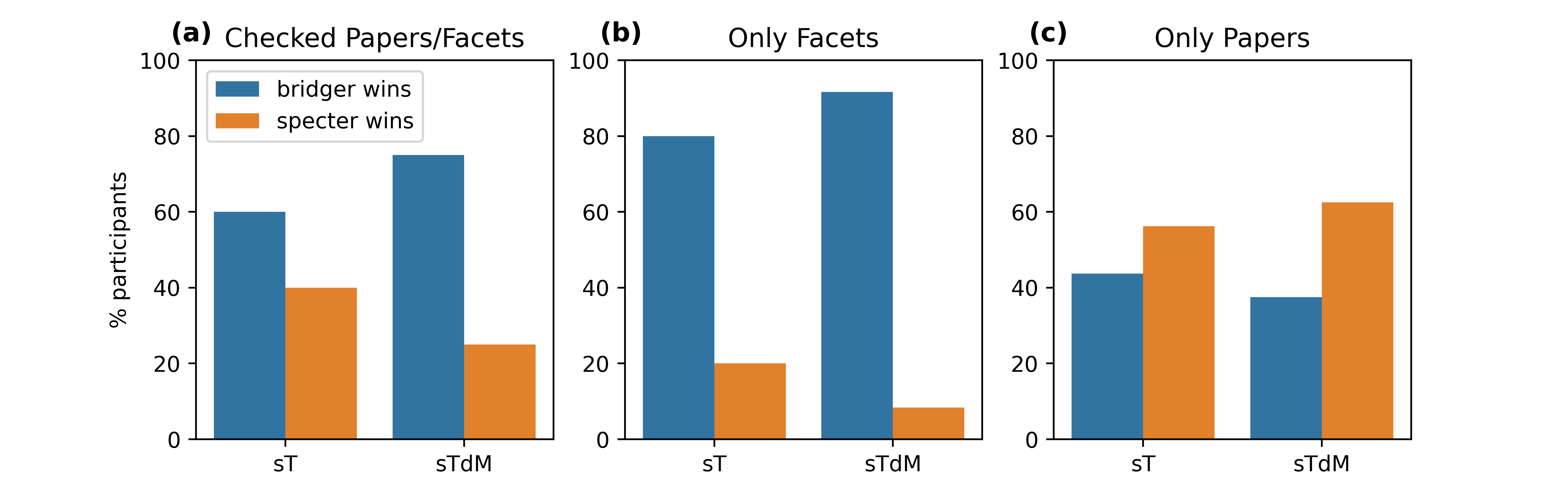}
  \caption{\emph{More users prefer Bridger for suggesting novel, interesting authors.} Percent of the participants who preferred author suggestions surfaced by faceted conditions (\simTask and \simTaskDistMethod, blue bars) compared to a baseline non-faceted paper embedding (\simspecter, orange bars). On average, users prefer the former suggestions, leading to more discovery of novel and valuable authors and their work (a). When broken down further, we find users substantially preferred the facet items shown for authors in our condition (b), and preferred the paper embedding baseline when evaluating papers (c). See §\ref{sec:exp2} for discussion.}
  \Description{(a) For checked papers and facets: for the similar-tasks condition, 60 percent of participants preferred Bridger author suggestions compared to 40 percent who preferred the specter author suggestions. For the similar-tasks-distant-methods condition, 78 percent preferred Bridger. (b) When considering checkboxes for only facets: for the similar-tasks condition, 80 percent of participants preferred Bridger over specter. For the similar-tasks-distant-methods condition, 96 percent preferred Bridger. (c) When considering checkboxes for only papers: for the similar-tasks condition, 41 percent of participants preferred Bridger over specter. For the similar-tasks-distant-methods condition, 38 percent preferred Bridger.}
  \label{fig:scores-basic}
\end{figure*}
%(a) For checked papers and facets: for the \simTask condition, 60 percent of participants preferred bridger author suggestions compared to 40 percent who preferred the specter author suggestions. For the \simTaskDistMethod condition, 78 percent preferred bridger. (b) When considering checkboxes for only facets: for the \simTask condition, 80 percent of participants preferred bridger over specter. For the \simTaskDistMethod condition, 96 percent preferred bridger. (c) When considering checkboxes for only papers: for the \simTask condition, 41 percent of participants preferred bridger over specter. For the \simTaskDistMethod condition, 38 percent preferred bridger.

For each author card evaluated by a user, we calculate the ratio of checked boxes to total boxes (typically 5-10 for papers, 10 for topics and facets; see §\ref{sec:exp2-design}) in that card. Then, for each user, we calculate the average of these ratios per condition (\simTask, \simTaskDistMethod, \simspecter), and calculate a user-level preference $\preferencescore$ specifying which of the three conditions received the highest average ratio. Using this score, we find the proportion of users who preferred each of the \simTask and \simTaskDistMethod conditions in comparison to \simspecter. This metric indicates the user's preference between Bridger- and \specter-recommended authors in terms of novelty and value (\textbf{RQ5}).

Figure \ref{fig:scores-basic}(a), shows results by this metric. The facet-based approaches lead to a boost over the non-faceted \simspecter approach, with users overall preferring suggestions coming from the facet-based conditions. This is despite comparing against an advanced baseline geared at relevance, to which users are naturally primed.

We break down the results further by slightly modifying the metric to account for the different types of items users could check off. In particular, we distinguish between the task/method/\allowbreak
resource/topic checkboxes, and the paper checkboxes. For each of these two groups, we compute $\preferencescore$ in the same way, ignoring all checkboxes that are not of that type (e.g., counting only papers). This breakdown reveals a more nuanced picture. For the task, method, resource and topic facets, the gap in favor of \simTask grows considerably (Figure~\ref{fig:scores-basic}b). In terms of papers only, \simspecter, which was trained on aggregate paper-level information, achieves a marginally better outcome compared to \simTask, with a slightly larger gap in comparison to \simTaskDistMethod (Figure~\ref{fig:scores-basic}c).
Aside from being trained on paper-level information, \specter also benefits from the fact that biases towards filter bubbles can be particularly strong with regard to papers. Unlike with facets, users must tease apart aspects of papers that are new and interesting to them versus aspects that are relevant but familiar.
See §\ref{sec:exp2qual} for more discussion and concrete examples.

Importantly, despite obtaining better results overall with the faceted approach, we stress that our goal in this paper is not to ``outperform'' \specter, but mostly to use it as a reference point~---~a non-faceted approach used in a real-world academic search and recommendation setting. We also note that \specter and other existing alternative baselines we could use are not tuned to our task of building bridges for authors across filter bubbles.

\paragraph{Personas}
We also compare the results from \simTask and \simTaskDistMethod conditions based on personas $\mathpersonas$ for user $\mathauthor$, versus the user's non-persona-based results presented above (\textbf{RQ6}).
We compare the set of authors found using personas with authors retrieved without splitting into personas (equivalent to the union of all personas). Table~\ref{tab:exp2:persona} shows the number of users for which the average proportion of checked items was higher for the persona-matched authors than for the overall-matched authors (for at least one of the personas). For most participants, users signalled preference for persona-matched authors when looking at one or both of their personas. Interestingly, for papers we see a substantial boost in preference for both conditions, indicating that by focusing on more refined \emph{slices} of the user's papers, we are able to gain better quality along this dimension too.

\begin{table}
\begin{tabular}{@{}lll@{}}
\toprule
Item type & sT & sTdM   \\ \midrule
Overall       & 58\%  & 75\% \\
Paper     & 83\% & 67\% \\
Topic     & 58\%  & 75\% \\
Task      & 42\%  & 50\% \\
Method    & 67\%  & 58\% \\
Resource  & 50\%  & 67\%
\end{tabular}
\caption{Percentage of users with personas (N$=$12), for which the average proportion of checked items was higher for the persona-matched authors than for the overall-matched authors. Users saw suggested authors based on two of their personas. The suggestions came from either the \simTask or \simTaskDistMethod conditions. Reported here are counts of users who showed preference for one or both personas. ``Overall'' shows results for all checkboxes considered in aggregate; this is followed by results for individual item categories (papers and facets).}
\label{tab:exp2:persona}
\end{table}

\subsection{Evidence of Bursting Bubbles}

\begin{figure}
  \centering
  \includegraphics[width=0.75\columnwidth]{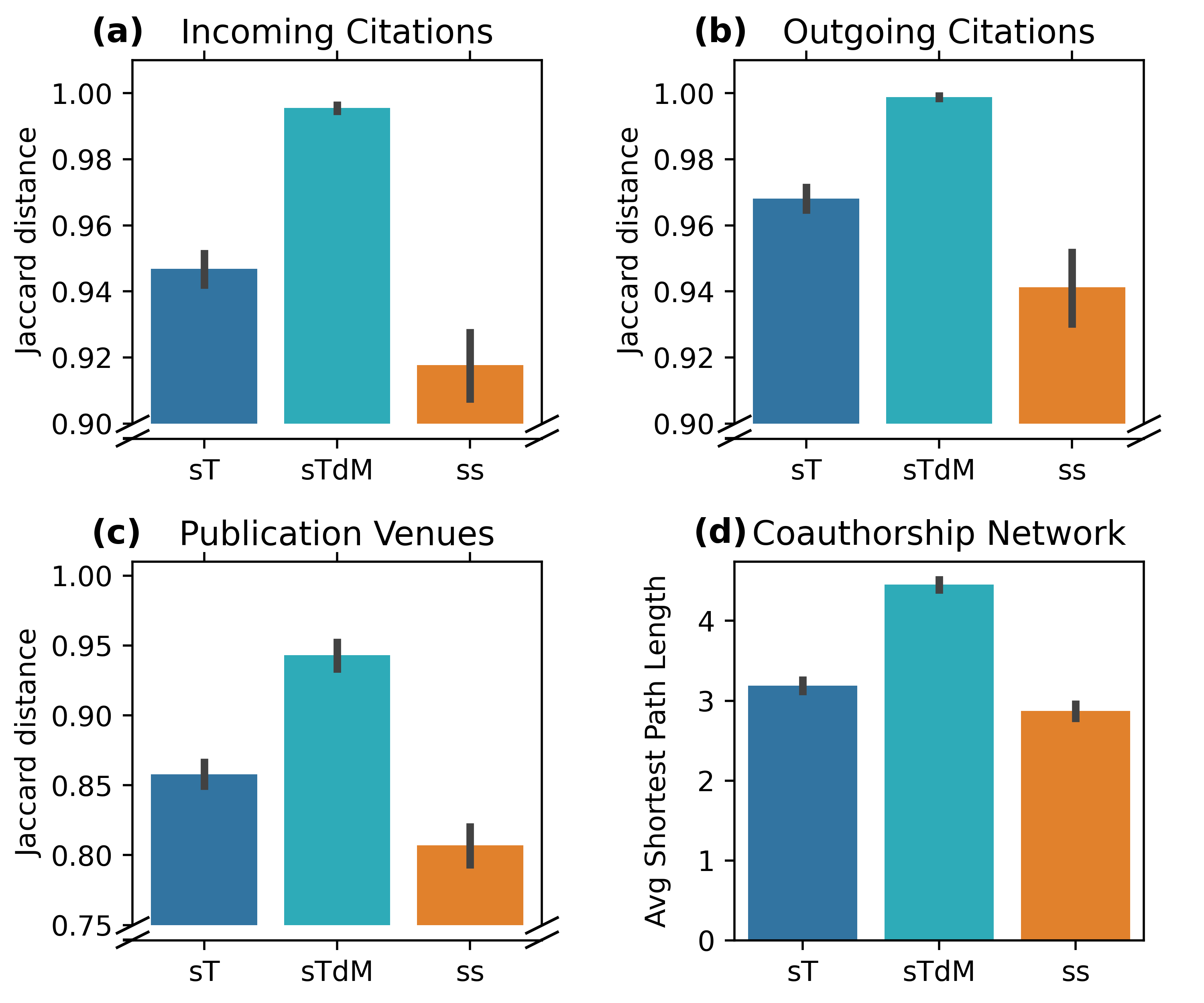}
  \caption{\emph{Bridger suggests authors that are more likely to bridge gaps between communities.} In comparison to the baseline, facet-based (Bridger) author suggestions link users to broader areas. Clockwise: (a, b) Jaccard distance between suggested authors' papers and the user's papers for incoming citations (a) and outgoing citations (b); greater distance means that suggested authors are less likely to be cited by or cite the same work. (c) Jaccard distance for publication venues. (d) Shortest path length in the coauthorship graph between author and user (higher is more distant). Bridger conditions (\simTask and, especially, \simTaskDistMethod) show higher measures of distance.}% Error bars are 90\% confidence intervals.}
  \Description{(a) Incoming citations Jaccard distances between suggested authors' papers and the user's papers: for similar-tasks: 0.95, for similar-tasks-distant-methods: 0.99, for specter: 0.92. Error bars representing 90 percent confidence intervals do not overlap. (b) Outgoing citations Jaccard distances between suggested authors' papers and the user's papers: for similar-tasks: 0.97, for similar-tasks-distant-methods: 1.0, for specter: 0.94. Error bars representing 90 percent confidence intervals do not overlap. (c) Publication venues Jaccard distances between suggested authors' papers and the user's papers: for similar-tasks: 0.85, for similar-tasks-distant-methods: 0.95, for specter: 0.81. Error bars representing 90 percent confidence intervals do not overlap. (d) Average shortest path length in the coauthorship network between the authors and the user: for similar-tasks: 3.1, for similar-tasks-distant-methods: 4.7, for specter: 2.9. Error bars representing 90 percent confidence intervals do not overlap.}
  \label{fig:other-distance-measures}
\end{figure}

The matched authors displayed to users were identified based either on \simTask and \simTaskDistMethod or the baseline \specter-based approach (\simspecter). These two groups differed from each other substantially according to several empirical measures of similarity. We explore the following measures, based on author dimensions in our data that we do not use as part of the experiment: (1) Citation distance: A measure of distance in terms of citations that the user has in common with the matched author (Jaccard distance: 1 minus intersection-over-union).
This is calculated both for incoming and outgoing citations. (2) Venue distance: The Jaccard distance between user and matched author for publication venues. (3) Coauthor shortest path: The shortest path length between the user and the matched author in the coauthorship graph.
Findings of this analysis, shown in Figure~\ref{fig:other-distance-measures}, suggest that Bridger surfaces novel authors from more diverse, distant fields and research communities than \specter (\textbf{RQ5}).

%(a) Incoming citations Jaccard distances between suggested authors' papers and the user's papers: for \simTask: 0.95, for \simTaskDistMethod: 0.99, for \simspecter: 0.92. Error bars representing 90 percent confidence intervals do not overlap. (b) Outgoing citations Jaccard distances between suggested authors' papers and the user's papers: for \simTask: 0.97, for \simTaskDistMethod: 1.0, for \simspecter: 0.94. Error bars representing 90 percent confidence intervals do not overlap. (c) Publication venues Jaccard distances between suggested authors' papers and the user's papers: for \simTask: 0.85, for \simTaskDistMethod: 0.95, for \simspecter: 0.81. Error bars representing 90 percent confidence intervals do not overlap. (d) Average shortest path length in the coauthorship network between the authors and the user: for \simTask: 3.1, for \simTaskDistMethod: 4.7, for \simspecter: 2.9. Error bars representing 90 percent confidence intervals do not overlap.

\subsection{Qualitative Findings: User Interviews}
\label{sec:exp2qual}
Experiment II's qualitative results were evaluated similarly to Experiment I (§\ref{sec:exp1}), using thematic analysis \cite{braun2012thematic}. The interviews support our quantitative results, affirming that Bridger authors encourage more diverse connections (\textbf{RQ5}). Under the Bridger conditions, participants noted diverse potentially useful research directions that connected their work to other authors not only within their own subareas, but also other areas. {This was especially true under the \simTaskDistMethod condition.} For instance, P9, who works on gradient descent for convex problems, saw a \simTaskDistMethod author's paper discussing gradient descent but for deep linear neural networks, which imply non-convex problems. They remarked, ``\emph{This is a new setup. It’s very different, and it’s super important \ldots %This is 
definitely something I would like to read \ldots}'' {Considering a paper under a \simTaskDistMethod author, P6 observed an interesting contrast with their work: ``\emph{I think my work has been bottom-up, so top-down would be an interesting approach to look at.}''} As another example, P2 drew a connection between the mathematical area of graph theory and their area of human-AI decision-making under the \simTaskDistMethod condition: ``\emph{This could be interesting mostly because \ldots they're using graph theory for decision making \ldots something I have not considered in the past.}''  P19 remarked of an \simTaskDistMethod author's paper, "\emph{This one actually seems quite interesting because it seems like explicitly about trying to bridge the gap between computational neuroscience models, understanding the neocortex, and computing. So that seems like it's... going to actually chart the path for me between my work and the stuff I think about like artificial neural networks and machines.}"

{In reacting to \simTaskDistMethod authors, many participants were able to go further than simply stating their interest in a connection and also describe \emph{how} they would utilize the connection. Looking at a \simTaskDistMethod author, P6 explained how the author's neuroscience work could motivate work in their area of natural language processing: ``\emph{I might learn from that [paper] how people compose words, and that might be inspiring for work on learning compositional representation \ldots''} P18 checked off a paper titled ``Multidisciplinary Collaboration to Facilitate Hypothesis Generation in Huntington’s Disease'' under a \simTaskDistMethod author ``\emph{because new ways to think about generating hypotheses could be interesting.}'' Seeing the topic `spike-timing-dependent plasticity' under a \simTaskDistMethod author, P19 mused, ``\emph{I would like to understand how spike-timing-dependent plasticity works and whether that could lead to a better learning rule for other types of neural nets, like the ones I work with on language, so that seems fun.}'' P12 described a \simTaskDistMethod author's paper about knowledge-driven search applications as useful to them because ``\emph{One of my primary research areas is knowledge base completion. However, that’s not an end application. An end application would be a search application which kind of uses my method to complete the knowledge base, and gives the user the end result. \ldots}''} Though the \simTaskDistMethod condition presented more of a risk in terms of surfacing authors with which the user could draw connections, it also surfaced the more far-reaching connections. 

{The \simTask condition also helped participants ponder new connections, though perhaps not as distant. P8 said of a \simTask author's work, ``\emph{I've worked a bit on summarization, so I want to know whether the approaches that I’ve worked on are applicable to real-time event summarization, which is a task I don't know about.}'' Also reflecting on a \simTask author, P1 explained, ``\emph{I’ve done a lot of work with micro tasks and these seem more maybe larger scale, like physical tasks or like planning travel. \ldots There are so many problems \ldots that I could apply my techniques to.}''} Other times, participants would connect one facet of their work to a different facet of the suggested author's work. In discussing a question-answering paper from a \simTask author, P8 explained, “\emph{I don't have experience with [the method] adversarial neural networks [used in this paper], but question answering is a task that I’ve worked on, so I would want to check this.}” 
Conversely, if participants found new connections with \specter, they tended to be more immediate connections to authors in their area. As an example, when checking off the paper ``Efficient Symmetric Norm Regression via Linear Sketching'' from a \specter-suggested author, P9 observed, ``\emph{I have used sketching techniques and I have [also] used norm regression, but [on] this specific problem I have not.}'' P9 also identified some of the papers from the suggested author as co-authored by their advisor. 

Although participants were asked to only check off interesting papers that suggested something new for them to explore, biases towards filter bubbles can be particularly strong with regard to papers because users must tease apart papers' new and interesting aspects from their relevant but familiar aspects. Even if a paper is directly connected to a user's research, they may be tempted to check off a paper because they have not seen that \emph{exact} paper or because it has minute differences from their work. For instance, P10 commented, "\emph{The problem is if it’s so general a title, I assume there’s something interesting happening, but I’m not completely sure.}” In contrast, when judging a particular facet item, participants need only contemplate the novelty of the term itself, without distraction or fixation on other terms~\cite{hope_accelerating_2017,kittur_scaling_2019, hope2021scaling}. As an example, P17 swiftly separated a task's general relevance from its lack of novelty to know not to check it. They explained, ``\emph{`Scientific article summarization'-  It is relevant, [but] I'm already familiar with it.}'' This bias helps explain the overall preference for \specter when considering only papers (Figure~\ref{fig:scores-basic}(c)).

\subsubsection{Personas} 
\label{sec:personas_qual} For the 12 participants who had personas, seven described their two personas as distinct, coherent identities that would be useful for filtering author suggestions. As an example, P2 characterized their personas as related to ``\emph{human-AI collaboration or decision-making}'' and ``\emph{error analysis and machine learning debugging}'' respectively.
The other 5 participants described one persona as coherent and seemingly useful for filtering authors. Concerns about their other personas were related to coherence, granularity, overlap with the other persona, and preference for the non-persona results after already looking through them and their first persona. 
Though the persona author suggestions performed relatively well in generating novel connections (Table \ref{tab:exp2:persona}), a few participants commented that they did not see the connection between suggested authors and their persona. For example, under a persona associated with lexical semantics, P6 commented on a \simTaskDistMethod paper, ``\emph{`Causality' is not one of the topics that I would work on in lexical semantics.}''

\section{Discussion, Limitations \& Design Implications}\label{sec:exp2interview}

We provide further analysis and discussion of our findings, including limitations of our proposed system and key challenges surfaced in the course of user interviews. We also discuss potential design implications for future author discovery systems. 

\subsection{Author Representations}\label{sec:exp2interview-context}
\label{sec:facetsNewButContext}

\subsubsection{Faceted Representation} Our experimental results point toward the advantages of a facet-based approach in the context of author discovery. We find that short, digestible items in the form of an author's tasks, methods, and resources can help participants consider interesting new research directions that they did not consider based on the author's papers alone. For instance, one participant (P14) expressed that a
Bridger-suggested author's paper associated with medical image diagnosis would not be useful for them to consider because ``\emph{breaking into that space for me would require a lot of work.}'' However, when they later saw `medical image diagnosis' as a task, they commented, ``\emph{As a task, I could see some usefulness there. There could be other approaches that might more quickly catch my interest.}'' Committing to interest in the overall task required much less effort. Moreover, participants were able to peruse more of an author's interesting tasks and methods that they did not necessarily find in their top papers. Reacting to one
Bridger-suggested author, P3 did not see any papers related to `biomedical question answering,' but they did see `biomedical question answering system' as a method. They then noted, ``\emph{I'm going to click `biomedical question answering' because that's not what I have worked on before, but I'm interested in learning about it.}'' 

Our work in the computer science domain made use of \textit{tasks, methods}, and \textit{resources} --- important functional aspects in this area \cite{luan_multi-task_2018}. Our results for Experiment I indicated that scientists find these more granular terms helpful both in describing their own work or the work of a known researcher, as well as for learning about unfamiliar researchers (§\ref{sec:exp1:rq1}, \ref{sec:exp1:rq3}). This suggests that to extend our approach more broadly to other areas of science, categories of terms that have important semantic meaning in specific domains are required, as opposed to using generic keywords or considering text in aggregate. In biomedical research, for example, salient facets might include the drugs that researchers study, or the diseases that their work addresses \cite{hope2020scisight}.
Alternatively, a future system could provide users with the ability to select or define which ``author dimensions'' matter to them, as opposed to assuming one universal pre-defined set. 

\subsubsection{Author Personas}

Our results in experiments I and II suggest that directly accounting for authors' different lines of work, or personas, can help boost user satisfaction in discovery systems. In our work we focused on a specific notion of personas based on clustering authors' papers, but this can be extended and generalized. For example, we could allow users to more directly select themselves subsets of papers for which they want to find interests, allowing users to define their own lines of work. This could also help with challenges in automatically clustering papers (§~\ref{sec:personas_qual}) by providing interactive feedback and supervision from users. Another important point for consideration is the temporal element, capturing author evolution over the years. Our current method simply looks 5 years back, but this could and should in principle be made adaptive per author, to account for different timelines for different authors. This too could potentially be made an interactive choice by the user, allowing them to segment their work temporally.

\subsection{Challenges with Novel Information \& Ideas}

\subsubsection{Novel Terminologies} An interesting design consideration that emerges in our experiments is that highly fine-grained terms (such as names of specific methods or datasets) can also introduce challenges in the context of discovery of \emph{novel, unfamiliar} authors and their work. In particular, as discussed in Experiment I, while overall our faceted representation of authors was considered more useful for understanding the work of new authors, unfamiliar names of methods also hindered understanding.

This tradeoff was also reflected in interviews in our user study evaluating author recommendations (Experiment II). Participants commonly identified tasks, methods, and resources as interesting, even when they did not fully understand their meaning. When P4 saw the method `least-general generalization of editing examples' from a Bridger-suggested
author, they stated, ``\emph{Don't know what this means exactly, but it sounds interesting.}'' P13 marked their interest in {the task} ``folksonomy-based recommender systems'' under a Bridger-suggested
author after commenting, ``\emph{I'm curious [about folksonomy] simply because I'm ignorant.}'' 
In seeing the resource `synaptic resources' under a Bridger-suggested
author,
P19 simply said, “\emph{I’d like to know what that is.}” Nonetheless, many participants also struggled with indiscernible terms. For example, P20 said of the resource `NAIST text corpus' under a Bridger-suggested
author, ``\emph{I'm not sure what this is, and I can’t guess from the name. And it wasn’t mentioned in the title of the papers.}''
P2 explained that a paper did not ``\emph{seem that interesting, but mostly because I don't understand all of these words.}'' 
Thus, providing term definitions may be helpful. {For additional context,} multiple participants expressed interest in having abstracts available, and P15 suggested including automated summaries~\cite{cachola2020tldr}.

This problem of ``unknown terms'' encountered by our participants increases the effort required from users, potentially deterring users from considering certain authors/directions. An important line of future work will be addressing this problem by providing just-in-time definitions of terms using extractive summarization~\cite{Narayan2018RankingSF} or generative approaches~\cite{Liu2019RoBERTaAR}. In particular, an especially appealing idea is to develop methods for \emph{personalized} explanations of new concepts, anchored in concepts with which the end user is already familiar (e.g., explaining a new neural network model by relating it to an older known one)~\cite{murthy2021personal}.

\subsubsection{Biases Toward Scientific Filter Bubbles}
\label{sec:biases}
An important challenge reflected in our results is that of time constraints in the fast-moving world of research, inhibiting exploration beyond the filter bubble. Despite clear interest in an author's distant research, a couple of participants in Experiment II were hesitant to make connections. For example, in reacting to a Bridger-suggested %\simTask 
author, P11 recognized, ``\emph{There's just a bunch of really interesting kind of theory application papers in this list that I'm not familiar with. \ldots I would maybe scan a little bit of these, but it's so far off that it's harder to make room to read someone that far away, but still cool.}'' 

Unknown background knowledge can make it difficult to consider new areas. Engaging with distant authors' work requires a large cognitive load that can impede uncovering connections. As P18 in Experiment II noted: ``\emph{Maybe there's some theoretical computer science algorithm that if I knew to apply it to my problem would speed things up or something like that, but I wouldn't know enough to recognize it as interesting.}'' This further suggests that unfamiliar terms can especially hinder making interesting connections, and that a personalized system design that highlights the most useful aspects of a distant author's research may facilitate building far-reaching connections. This also further compounds aversion to novel ideas, and fixation on familiar frames of problems and solutions \cite{fu_meaning_2013,hope_accelerating_2017,chan2018solvent,kittur_scaling_2019, hope2021scaling}. Because Bridger's authors are selected to be more distant from the user than \specter's authors, they sometimes met with hard-line resistance, without full consideration of potential links. Looking at a Bridger-suggested
author, the natural language processing (NLP) researcher P20 said, ``\emph{This is not really an NLP paper, so I would pass.}''
Similarly, P17 rejected a paper from a
Bridger suggestion, saying ``\emph{I don't know anything about neuroscience, and I'm not going to start now probably.}''

\subsection{Data \& System}

One limitation of our work is that many early-career researchers are excluded from our system because they do not have enough papers. This user group, however, could potentially especially benefit from this type of discovery system. This problem relates more broadly to the much-discussed ``cold start'' challenge in recommendation systems~\cite{bobadilla_collaborative_2012,lam_addressing_2008}. One potential implication for future systems is to provide such users with alternative options, such as viewing recommendations to authors they consider relevant (e.g., advisors, mentors, etc.).

Another limitation lies in the accuracy of the information extraction methods we employ. This is an issue that impacts all work that relies on such methods; however, it is somewhat mitigated in our setting by aggregating over many spans extracted from authors' papers, which averages out some noise. Additionally, other work has found that even with moderate extraction accuracy, strong ideation utility can be obtained~\cite{hope2021scaling}. The problem of identifying connections between mentions of tasks, methods or resources across scientific papers is very much an open one \cite{cattan2021scico}; as future models in this area become more precise, our approach for matching authors is expected to become more accurate, too.

Finally, our evaluation focused primarily on surfacing authors who spark new ideas outside users' familiar areas. We design controlled studies measuring our approach's ability to surface inspirations in comparison to a real-world baseline scientific search model. An interesting and challenging direction for further evaluation is to measure Bridger's longer-term ability to provide useful inspirations that yield viable research directions and projects, and to measure the system's ability to help users in less controlled settings. This type of evaluation would require longer-term interaction with the user, and longitudinal observations.

\section{Conclusion}

We presented Bridger, a framework for facilitating discovery of novel and valuable scholars and their work. Bridger consists of a faceted author representation, allowing users to see authors who match them along certain dimensions (e.g., tasks) but not others. Bridger also provides ``slices'' of a user's papers, enabling them to find authors who match the user only on a subset of their papers, and only on certain facets within those papers. Our experiments with computer science researchers show that the facet-based approach was able to help users discover authors with work that is considered more interesting and novel, substantially more than a relevance-focused baseline representing state-of-art retrieval of scientific papers. Importantly, we show that authors surfaced by Bridger are indeed from more distant communities in terms of publication venues, citation links and co-authorship social ties. While our work only considers the domain of computer science research, we believe the techniques could generalize outside of computer science, potentially connecting people with ideas from even more disparate fields as we make steps toward bridging gaps across all of science. These results suggest a new and potentially promising avenue for mitigating the problem of isolated silos in science.

% \section{SIGCHI Extended Abstracts}

% The ``\verb|sigchi-a|'' template style (available only in \LaTeX\ and
% not in Word) produces a landscape-orientation formatted article, with
% a wide left margin. Three environments are available for use with the
% ``\verb|sigchi-a|'' template style, and produce formatted output in
% the margin:
% \begin{itemize}
% \item {\verb|sidebar|}:  Place formatted text in the margin.
% \item {\verb|marginfigure|}: Place a figure in the margin.
% \item {\verb|margintable|}: Place a table in the margin.
% \end{itemize}

\begin{acks}
We thank Kishore Vasan, Jonathan Borchardt, and Jonathan Bragg for their contributions to this project. We also thank our study participants, and the four anonymous reviewers for their helpful suggestions. This work is partially supported by NSF Grant OIA-2033558, NSF RAPID grant 2040196, and ONR grant N00014- 18-1-2193.
\end{acks}

\newpage
\bibliographystyle{ACM-Reference-Format}
\bibliography{scisight}

% \appendix

\end{document}